%% file: Manoury_etal_arxiv.tex
\documentclass{article}
\usepackage{arxiv}

\usepackage[utf8]{inputenc} 
\usepackage[T1]{fontenc}    

\usepackage[bookmarks=false]{hyperref}
    \hypersetup{colorlinks,
      linkcolor=blue,
      citecolor=blue,
      urlcolor=blue}
\usepackage{svg}
\usepackage{pdfpages}
\usepackage{enumitem}
\usepackage{booktabs}

\newcommand{\colrule}[0]{\hline}
\newcommand{\botrule}[0]{\hline}

\usepackage[acronym]{glossaries}
\input{glossaries.tex}
\setacronymstyle{long-short}
\makeglossaries

\usepackage{tikz}
\usepackage{multirow}

\title{Supporting Changes in Digital Ownership and Data Sovereignty Across the Automotive Value Chain with Catena-X}

\author{ \href{https://orcid.org/0000-0002-3443-4937}{\includegraphics[scale=0.06]{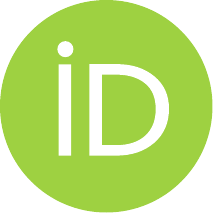}\hspace{1mm}Marvin Manoury}\\
  Fraunhofer Institute for \\Production Systems and Design Technology IPK\\
  Pascalstrasse 8-9\\ 10587 Berlin, Germany\\
	\And
	Theresa Riedelsheimer \\
  Fraunhofer Institute for \\Production Systems and Design Technology IPK\\
  Pascalstrasse 8-9\\ 10587 Berlin, Germany\\
	\AND
  \href{https://orcid.org/0000-0002-2095-662X}{\includegraphics[scale=0.06]{orcid.pdf}\hspace{1mm}Malte Hellmeier}\\
   Fraunhofer Institute for \\Software and Systems Engineering ISST\\
   Speicherstraße 6 44147 Dortmund, Germany\\
   \AND
   \href{https://orcid.org/0009-0009-4625-8561}{\includegraphics[scale=0.06]{orcid.pdf}\hspace{1mm}Tom Meyer} \\
   Fraunhofer Institute for \\Software and Systems Engineering ISST\\
   Speicherstraße 6 44147 Dortmund, Germany}

   \date{}
   \renewcommand{\headeright}{}

\begin{document}
\maketitle

\begin{abstract}
	\input{abstract.tex}
\end{abstract}

\keywords{Digital Twins \and Ownership \and Circular Economy \and Data Sovereignty \and Supply Chain \and Catena-X}

\section{Introduction}
\label{sec:intro}

Most modern functionalities of products, associated services, or businesses require accurate, up-to-date information on that product. 
The concept of \glspl{DT} is an approach for storing and making this information accessible and fulfilling some intelligent functionalities.
There exists a plethora of definitions for a \gls{DT}. 
Some examples that are considered as baseline in this article are Stark \& Damerau \cite{Stark2019}, Grieves \cite{Grieves2018, Grieves2022, Grieves2016} and Catena-X \cite{CatenaX2024} also presented in Mügge et al. \cite{Muegge2022, Muegge2023} which will be used as an example.

The most important aspect of the definitions is that a DT consolidates data on an individual asset based on the common product type and instance data that is generated and updated across the lifecycle of that asset, including possible business information in the production development and manufacturing phase as well as information from usage, maintenance, and dismantling. 
New regulations from the EU, for example, within the \gls{ESPR} and the Digital Product Passport \cite{Environment30.03.2022} that is mandatory for specific batteries from 2027 \cite{EPCEU12.07.2023} as well as the expected circular vehicle passport \cite{EuropeanCommission2020} drive automotive companies to collect and provide more specific data on their vehicles, components, and materials -- even across lifecycles, as the example of secondary material content shows. 
Therefore, especially within the context of \gls{CE}, the phases after dismantling are also to be considered. 
Hence extending \gls{DT} data with information on reuse, remanufacturing, refurbishment, repurposing, recycling, or recovery \cite{Muegge2022, Muegge2023,Potting2017}  is an essential part of success.

This paper presents the results from a specific application in the automotive industry based on one of the most significant German digitalization research projects, Catena-X \cite{CatenaXAutomotiveNetworke.V..}. 
After providing a basic background of the research project Catena-X and a short example, section 3 describes the approach for generating new concepts and functionalities within the project. 
In the results and discussion sections, approaches for updating DT are described and evaluated.

In Catena-X, the concept of the Asset \gls{AAS} \cite{IndustrialDigitalTwinAssociation.June2023, IndustrialDigitalTwinAssociation.March2023} is used for the realization of the \glspl{DT}, with \textit{aspects} or \textit{submodels} used as structuring elements to aggregate relevant information. 
These submodels can be created and standardized by anybody with an appropriate use case. 
To enable interoperability, a common framework and ruleset called \textit{industry core} has been made that governs the harmonization of existing aspects and the core elements for aspects, the definition of an ID, the categorization of DTs with regard to instances and types, as well as the lifecycle phase and status of the \gls{DT}. 
Detailed information can be seen in the documentation of the open-source Tractus-X project \cite{TractusX2024}, associated with Catena-X.

This industry core also handles the creation of \glspl{DT}. 
In Catena-X, which mainly focuses on the automotive industry and thus has a deep OEM-supplier relation in the creation of the vehicle, initially, the manufacturer creates the DT to preserve control over it, the guiding principle behind data sovereignty \cite{Hellmeier2023,Jarke2019, Scherenberg2024} meaning self-determination over the use of data. 
In multi-sourcing scenarios, a customer needs to create one DT of each asset per supplier for two reasons: (I) competing suppliers must note synchronize their identifier (Catena-X ID) due to antitrust law requirements, and (II) after completing the sourcing of the a component at different vendors, the customer states that the given parts, which might be different in their composition, fulfill the customer's specification. 
The Catena-X ID and a company's business partner identification are used in aspects to build data chains.

As soon as the asset is produced, further stakeholders besides OEM and suppliers become relevant, including repair shops, dismantlers, and second-life users. 
Updating across the life cycle becomes challenging when a separate \gls{DT} is created for each of them. 
Also, data ownership must be considered as the basis for data sovereignty. 
In Catena-X, the separation between the data owner and data provider follows the IDS-RAM~\cite{Otto.2022}. 
The IDS-RAM is the Reference Architecture Model for Data Spaces governed by the International Data Spaces Association and differentiates ownership into possession and property, which aligns with the standard legal understanding. 
This means that, e.g., one party may be the data owner, but another may provide or use the data under specific conditions \cite{Otto.2022}. 
This paper focuses on the results generated in Catena-X for updating \glspl{DT} along an assets lifecycle and its influence on data ownership, data sovereignty \cite{Hellmeier2023,Jarke2019, Scherenberg2024}, and digital continuity \cite{Manoury2023,NationalArchives}.

\section{Example}
\label{sec:example}

An example scenario from the automotive industry shall be used to demonstrate the challenges and proposed solutions. 
The main relevant phases are the market research or predesign, the design, the specification, the product planning, the individual order and production start, production, supply and delivery, the usage phase with maintenance and repair, as well as the \gls{EoL} phase with its sub-phases and different \gls{CE}-strategies. 
For this paper, as a specific example, a vehicle is designed after market research by an OEM, and components are specified for in-house production but also their suppliers. 
Three suppliers (suppliers A, B, and C) provide different components as determined by the OEM, which can be seen as the status \textit{as delivered}. 
The vehicle is manufactured, e.g., assembled, by the OEM (\textit{as built}) and then sold in a \gls*{B2C} context to a consumer using it for 20 years before it reaches its' end of life (\textit{as used}). 
The estimated average lifespan in Western European Countries is 18.1 years \cite{Held.2021}. 
A repair is included to showcase a change in vehicle-component relationship. 
In the usage phase, the gearbox produced by supplier A and the battery with cells of supplier C are exchanged at different times. 
The vehicle's battery is exchanged at an OEM-certified repair shop, and the gearbox is exchanged at an independent workshop. 
At both times, at least the \gls{BoM} of the vehicle and the status of the exchanged components need to be updated and made accessible to each relevant participant in the network while preserving the data sovereignty of the data owners. 
Dynamic data, such as the mileage of the components, is also applicable across the whole lifecycle of the vehicle. 
Ultimately, the car is dismantled (\textit{as dismantled}) at an authorized company, some elements are reused, refurbished, or remanufactured, and the value material is recycled. 
As soon as the car enters the dismantling phase, it needs to be digitally registered that it reached the \gls{EoL}, and the statuses of the components need to be updated, including what decisions are taken about CE, meaning if they are refurbished or reused in some other context.

All these actions require updates of the vehicle's \glspl{DT} and components from different entities and shall be considered for evaluating the approaches in this use case. 
Considering the basic setup mentioned in the introduction, the scenario has to be regarded as depicted in Fig. ~\ref{fig:DTcreationscenario}.
The diagram uses a non-exhaustive form to ease understanding. 
This scenario shall be used as the foundation for the explanation of further investigations.

\begin{figure}[h]
	\centering
  \resizebox{0.9\columnwidth}{!}{\input{fig1_DTcreation.tex}}	\caption{Sample figure caption.}
	\caption{Digital Twin creation scenario over assets lifetime with Supplier, OEM, Consumer, Repair Shop and Dismantler as relevant stakeholders}
  \label{fig:DTcreationscenario}
\end{figure}

\section{Material and Methods}
\label{sec:material}

\begin{figure}[b]
  \centering
    \includegraphics[width=0.9\textwidth]{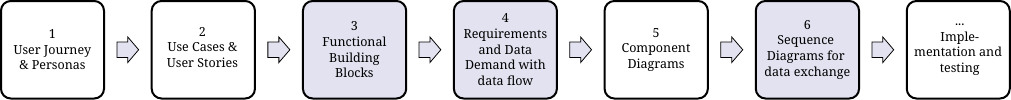}
  \caption{Approach to identify requirements and solutions}
  \label{fig:approach}
  \end{figure}

In the \gls{CE} use case in the Catena-X project, a V-Model \cite{VDI/VDE2021} oriented approach as an abstraction of Riedelsheimer et al. \cite{Muegge2023, Riedelsheimer.b, Riedelsheimer.2021} has been used for the user-centered design of DTs and required components. 
A customer journey was developed in the first phase, and the primary personas and their basic needs were identified. 
On this basis, relevant stakeholders, including dismantlers, OEMs, and suppliers, have defined persona-specific use cases and user stories.
Each user story has been analyzed for so-called \gls{FBB}, meaning a combination of basic functionalities to fulfill these user stories.
Examples of these functional building blocks are \textit{Finding DTs for a specific \gls{VIN}}, \textit{Requesting information on that vehicle}, and \textit{Updating information on that vehicle}. 
For each \gls{FBB}, requirements, and data demands, including the respective data flow and providers, have been aggregated. 
The data providers are the ones responsible for defining the usage and access policies for the data access.  
Next, relevant or affected IT components for the realization within Catena-X have been identified. 
These components have been classified as either central components that must be delivered by the Catena-X network or their implementation requirements for teams realizing relevant use case-specific applications. 
The relevant IT components have then been aggregated in component diagrams, and for the implementation of the data sharing, sequence diagrams have been created. 
As an example, the EcoPass-KIT can be investigated \cite{TractusX}. 
The specification process is visualized in Fig.~\ref{fig:approach}, and the main focus of this paper is highlighted. 
Eventually, the implementation and testing cycles were initiated.

Catena-X overall used the \gls{SAFe} as a process model to organize the different use cases and platform teams into Agile Release Trains \cite{SAF14.03.2023}.
The different teams' deliverables were separated into platform and business capabilities on the enterprise level. 
Architecture roles occurred on the team, use case, and enterprise level. 
While requirements for implementation teams have been evaluated directly with relevant stakeholders and application developers of \gls{COTS} and \gls{FOSS} solutions on the first two levels, the central component was discussed on the enterprise architect level.
Besides the differentiation of the teams, special interest groups have been built to provide governance on, e.g., creating data models, driving the platform architecture, sovereign data sharing, and DevSecOps practices. 
With them, the platform architecture has been considered, and possible solutions have been evaluated. 
The results were documented as \glspl{ADR} supported by sequence diagrams to show the required interactions on a technical level. 
The implementation is currently still under development (as of May 2024).

\section{Results}
\label{sec:results}

With the DT-V-Model approach mentioned in the previous section, in addition to general \gls{CE} requirements, such as \textit{organization-related certificates} (e.g. for authorized dismantling companies), \textit{reporting of regulatory information} \cite{EPCEU12.07.2023}, \textit{supporting generic input values as search requests} (e.g., VIN) and \textit{access control}, the \gls{FBB} \textit{updating DTs} has been identified. 
The main (clustered) functional (f) and non-functional (nf) requirements that were identified within workshops and interviews with the respective personas along the lifecycle are depicted in Table 1. The relevant components that have been identified to fulfill these requirements are:

\begin{table}[h]
  \caption{Requirements for Updating Digital Twins}
  \begin{tabular*}{\hsize}{@{\extracolsep{\fill}}p{3.5cm}p{11.5cm}c@{}}
  \toprule
   & Description & Type\\
  \colrule
  Data flagging	 &
  As an asset owner / operator, I want to be able to set and review the status of a DT (e.g. "has been reused", "has been sold", "has been maintained", "has been dismantled", “has been remanufactured”, “has been recycled”, “has been transferred to waste”).	&  (f)\\
  Certify DT status update with asset-specific certificates for major status change	  & 
  As a dismantling lead, I want to update and certify the status of a vehicle that will be dismantled and create a "Certificate of Decommissioning" ("Verwertungs-Nachweis") to do so. This also applies to certificates for re-used, remanufactured, refurbished, repaired, recycled, or labeled as waste.	& (f)\\
  Update DT with maintenance / usage information	 &
  As a workshop operator / operator of vehicle fleets / reseller, I want to update the DT information with data on usage (e.g. accidents) and maintenance or repair, specifically exchanging components (“detachment” and “re-attachment process").	 &  (f)\\
  Update DT information with dismantling results	&
  As a dismantling lead, I want to be able to update the component-vehicle relationship after the dismantling "detachment process."	& (f) \\
  Update DT information with dismantling results	&
  As a dismantling lead, I want to update which strategy was chosen for specific components.	& (f) \\
  Update DT with remanufacturing results	&
  As a remanufacturing lead, I want to update the information that certain components with specific IDs are integrated into manufacturing a new vehicle with a specific VIN (“re-attachment process").	& (f) \\
  Update DT with recycling results	&
  As a recycling lead, I want to update the information that certain components are recycled, to which material they are processed, and to report the recycling quota. & (f) \\	
  Requesting updated DT data	&
  As an OEM / manufacturer, I want to be able to access information on the actual dismantling results of vehicles and their components to execute design optimization for future product generations.	& (f) \\
  Requesting updated DT data	&
  As an OEM / manufacturer or regulatory body, I want to be able to access the quota of reused components or recycled materials after dismantling to support reporting for regulatory reasons.	& (f) \\
  Requesting updated DT data	&
  As a manufacturer, I want to access the actual secondary material content of material that I use within manufacturing.	& (f) \\
  Permissions/Ownership	&
  There has to be a possibility of updating the DT of assets in the field without being legally contracted to the OEM that created the "as-built" twin of the vehicle, e.g., for repair shops.	& (f) \\
  Historical data	&
  As a DT data consumer, I want to be able to access the history and different versions of the DT and the asset (e.g., the mileage must not be reset as soon as a component is included in a new vehicle) for at least the calculated lifetime of the respective asset or as regulatorily demanded \cite{EuropeanCommission2020,Environment30.03.2022}.	& (nf) \\
  Source of information	&
  When searching for a DT, the most recent information needs to be identifiable.	& (nf) \\
  Loss of data provider	&
  When a company is no longer part of Catena-X, the data (e.g., as handled for bankruptcy) must be transferable to relevant other parties.	& (nf) \\
  Scalability	&
  The solution should support prospective support of DT updates to relevant events, such as maintenance, for the registered vehicles (50 million passenger cars in 2024 in Germany \cite{KraftfahrtBundesamt.2024}).	& (nf) \\
  Scalability &
  The solution should support prospective DT updates to relevant events, such as vehicle dismantling, for the number of vehicles dismantled in Germany at authorized dismantling facilities (~400,000 per year in 2021 \cite{UmweltbundesamtUBA.2024}).	& (nf) \\
  \botrule
  \end{tabular*}
  \label{tab:requirements}
  \end{table}

\begin{itemize}
\item components making the DTs retrievable, which are in the case of Catena-X, the Digital Twin Registries \cite{TractusX27.04.2024a} implementing a registry and discovery interface in compliance with the AAS specification v3.0 \cite{IndustrialDigitalTwinAssociation.June2023,IndustrialDigitalTwinAssociation.March2023}, 
\item components for identifying oneself, e.g., as dismantler \cite{FOIS},
\item components for definition and negotiation of the conditions for the data identification and sharing (see “Connector") \cite{TractusX27.04.2024a} and
\item components for hosting the actual information that need to be implemented in \gls{AAS} specification v3.0 for submodel interfaces according to the Catena-X profile \cite{IndustrialDigitalTwinAssociation.June2023, TractusX27.04.2024a}
\end{itemize}

As all these components are fully central or decentralized components within the network that are not solely relevant to the use case, the functionality was investigated with central architects. 
Three main approaches to updating shared \glspl{DT} have been identified in the discussions. 
All three approaches focus on the OEM-supplier collaboration network mentioned before. 
Haße et al.~\cite{Hae.2022} characterize shared \glspl{DT} based on the aim to integrate data across company borders and the asset life cycle phases in distributed systems, allowing to “[\ldots] enrich the shared Digital Twin with data from external organizations” \cite[pp. 754 ff.]{Hae.2022}. 
The approaches considered within the project are:

\begin{enumerate}[label=(\arabic*)]
\item \textit{One Digital Twin}, 
with read and write access privileges for both the Twin-OEM and third parties; 
\item \textit{Several Digital Twins with read rights}
 on each copy of the DT at the stakeholder for the Twin-OEM and; 
\item \textit{Several Digital Twins with licensing and notifications},
 where stakeholders can read the respective information from the OEM-twin, and notification mechanisms are established to integrate updates into the OEM-twin.
\end{enumerate}

In the following, each approach is described and presented in detail.

\subsection{One Digital Twin (Approach 1)}

\begin{figure}
  \centering
  \resizebox{0.7\columnwidth}{!}{\input{fig3_approach1.tex}}
  \caption{Update Scenario 1 - One Digital Twin}
  \label{fig:approach1}
\end{figure}
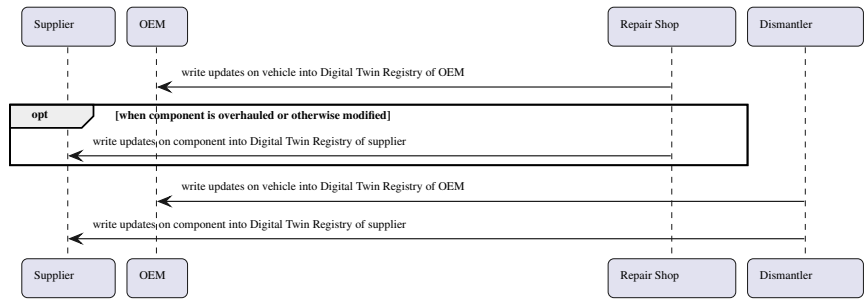

The first approach uses one DT per asset, which is located and stored at the OEM for a vehicle and at their suppliers for the components used in the vehicle (see Fig.~\ref{fig:approach1}). 
These \glspl{DT} have read/write access from all involved parties either through direct access to a Digital Twin Registry or a portal that updates the \gls{DT}. 
Considering the scenario mentioned in section~\ref{sec:example}, this means updates are directly written into the \gls{DT} located at the OEM or the supplier. 
All stakeholders need to have full control of these \glspl{DT}.

\subsection{Several Digital Twins (Approach 2)}

\begin{figure}[b]\vspace*{4pt}
  \centering
  \resizebox{0.8\columnwidth}{!}{\input{fig4_approach2.tex}}
  \caption{Update Scenario 2 - Several Digital Twins}
  \label{fig:approach2}
\end{figure}
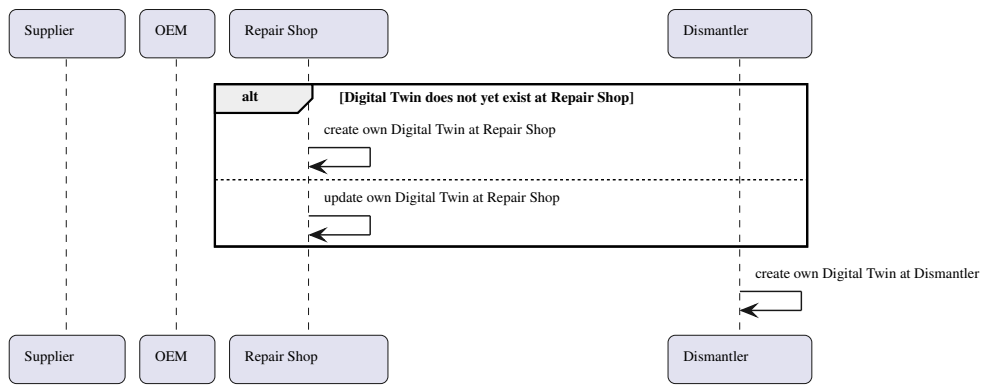

In the second approach (see Fig.~\ref{fig:approach2}), all stakeholders have \glspl{DT} that contain their specific information on the asset. 
This means that as soon as the repair shop gets the component and exchanges some components, the repair shop has to create another DT of the vehicle and the component that only includes the information generated through the repair. 
In this way, all stakeholders can define their own policies for their data to achieve data sovereignty. 
Other stakeholders only get read access to the DT information to aggregate all information on the vehicle they can access. 
If the policies do not allow copying, the data must be aggregated from various sources to access a complete overview of information. 
Thus, a common identifier like the \gls{VIN} for a vehicle must be used to link all information on the asset.

\subsection{Several Digital Twins with Licensing and Notification (Approach 3)}

\begin{figure}[t]\vspace*{4pt}
  \centering
    \resizebox{0.8\columnwidth}{!}{\input{fig5_approach3.tex}}
  \caption{Update Scenario 3 - Several Digital Twins with Licensing and Notification}
  \label{fig:approach3}
\end{figure}
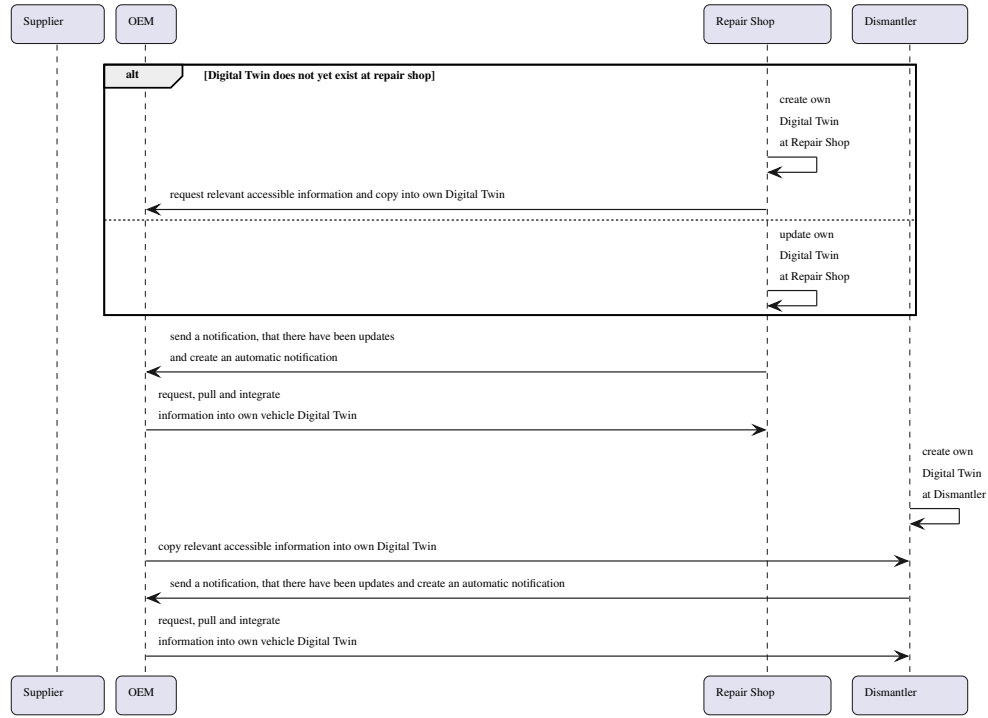

As in scenario 2, in this scenario, all stakeholders have their own DTs in the licensing and notification approach (see Fig.~\ref{fig:approach3}). 
The main difference is that as soon as a new \gls{DT} is created, there is a mandatory policy to copy the relevant information through a notification approach into the first \gls{DT}. 
That means that each partner (e.g., repair shop, dismantler) who wants to update information on an asset creates a copy of the mandatory information of the respective \gls{DT} if it does not yet exist at the partner and integrates the updates into their \gls{DT} as in scenario 2. 
The relevant information in this update has to have a policy that allows the original DT creator to read and copy the data. 
Additionally, a notification mechanism is set up to notify the original DT creator of asset data updates. Each data-sharing step (copy the twin, notify the twin creator, integrate information) needs to consider the licensing policies. 
Relevant information that has to have a copy policy might be the mileage of a component, the current BoM, or the state of health of a battery. 
Information not specified as mandatory can be provided by updating stakeholders on their own policies. This means the original DT, e.g., at the OEM, includes all relevant information from the original DT creator that has been copied from the updating parties. In contrast, more information can be accessed via the other stakeholders' DTs.

\section{Discussion}
\label{discussion}

All three approaches have different influences on data sovereignty, digital consistency, and ownership of the DT, discussed in the following. 
It has to be mentioned that all approaches and this discussion only focus on the technical realization, not on required contracting, etc.

While approach 1 (one DT) is theoretically very easy, the core challenge is access control from external parties. 
As the data is hosted at the OEM or supplier, there is only a single point of contact for all information, which supports digital continuity. 
But to keep data sovereignty for the data owners, the OEM or supplier hosting the data and the data provider in the sense of an updating party (e.g., repair shop), must be able to define how the data can be accessed and used. 
This can lead to conflicts, e.g., if a repair shop wants to grant access to some repair data to enable new business models, but the supplier providing the data of the DT prevents data sharing with other external suppliers. 
This must be handled with bilateral agreements per information input, which can be very difficult. 
Further, access control must be integrated on the data set level as otherwise, the updating parties can interfere. 
Overall, this approach's fulfillment of data sovereignty is doubtful because no self-determination is maintained due to the external data storage at one location. 
Besides this, the general ownership is also unclear: Does the OEM / supplier hosting the DT own all data in it or only the shell aggregating the data? What happens once the asset is sold, and the DT is sold as an asset as well, or in a case of bankruptcy? Who gets licenses on which information? 
Especially considering ownership, this approach lacks some further details. 
As all data is shared in this approach, access management and ownership in an OEM-supplier and third-party collaboration is very complex. 
Therefore, this approach has not been further investigated in Catena-X.

The ownership in approach 2 (several DT) is no issue, as all stakeholders have their own DTs and thus own the information they created with the update. 
This supports data sovereignty, as all stakeholders can define their own policies and manage access to their part of their DT data. 
With this possibility of managing access and information, the stakeholders must also keep the data current. 
This includes updating the DT whenever required, e.g., once the component is set as EoL state at the dismantler. 
The challenge in this approach is information is scattered across different stakeholders. A clear linking of information is required, e.g., by a central network service that discovers all stakeholders for a \gls{VIN}. 
In Catena-X, this is handled through discovery services \cite{TractusX27.04.2024a}. This enhanced data sovereignty solution raises data traffic, as the decentralized storage requires more calls to access data. 
The main challenge is to evaluate and aggregate the most recent information for one purpose. 
This is handled on an application level, not with a central service. For most use cases, this approach is currently used within Catena-X.

The third approach (several DTs with licensing and notification) has a single information point, as in approach 1. 
In this way, digital continuity is supported as any stakeholder or service has a clear location for information retrieval. 
Opposed to approach 1, the DT has no write permissions for all stakeholders but only read permission for other stakeholders with policies for access to only specific parts of the DT, e.g., data relevant to repair operations. This leads to better data sovereignty compared to approach 1 but still lacks detailed control mechanisms due to limited read and write access. As in approach 2, the updating stakeholders create themselves a DT with their own information. Additionally, they must create a policy allowing the original DT creator to copy/store and share the information. In that way, the original DT creator has all the mandatory information, and the updating parties can still offer the data with their own policies. The biggest difficulties here are handling licensing and flagging external data. The DT creator must be aware of the usage policies of the data provided by the updating stakeholders and apply them in their own DT, e.g., by copying the policies with the data into his or her DT. However, liability must also be defined correctly by flagging information as copied from external parties, as it can be difficult to verify all information. Also, there is a practical need for this data to be shared with other partners, such as in Fig.~\ref{fig:approach2} and Fig.~\ref{fig:approach3}, where the dismantler must be allowed to see the data provided by the repair shop. To scale this approach, the linkage of data and the component information must be available to all potentially involved stakeholders. Special information, such as specifications, may rely on different usage conditions. Table 2 summarizes the presented points.

\begin{table}
  \caption{Identified potentials of approaches}
  \begin{tabular*}{\hsize}{@{\extracolsep{\fill}}cp{4.2cm}p{4.2cm}p{4.2cm}@{}}
  \toprule
   & (1) Shared Digital Twin & (2) Several Digital Twins & (3) Licensing and Notification  \\
  \colrule
  Digital Consistency &   
  ++ single point of contact for all information for an asset for all consumers  &
  - several sources for information for an asset &
  ++ single point of contact for all validated and trustworthy information of an asset\\
  Data Sovereignty  & 
  \-\- no data sovereignty, limited policies, central storage & 
  + data sovereignty, as everyone can define their own policies, decentralized storage
   &
   + some data sovereignty, with limited read and write permissions\\
  Ownership &  
  \-\- unclear license and ownership situation &  
  ++ clear ownership, no special licenses needed &
  +  clear ownership, but licenses needed \\
  Level of sharing &  
  Sharing all data with the Twin-OEM &
  Sharing responsibility to link DTs &
  Share updates and keep own DT in sync \\
  \botrule
  \multicolumn{4}{l}{\-\- = not fulfilled, - = insufficient, + fulfilled to some extend, ++ fulfilled}
  \end{tabular*}
  \end{table}
  
  When specifically considering ownership, in the first approach, once the ownership changes, e.g., because a component is sold for a non-automotive use case and the initial component producer is no longer liable for the information, the DT might no longer be hosted and owned by the original creator. All services built on this information must change the asset's endpoint. This might require a forwarding approach from the original DT creator to the new owner. Also, the bankruptcy of stakeholders or if they are leaving the environment is a big issue, as all data must be transferred somewhere else; otherwise, it is lost to the network and services built on it. Approach 2 and 3 somehow lower the risk as the data is duplicated across different stakeholders. In approach 3, it is still open to clarifying how the copied data is handled in a case of ownership change. License agreements need to be made to define where sharing of information ends. The linking of information that is not copied to the originally created DT is also an issue that needs to be solved. This is also relevant from a liability point: Somehow, it must be shown that data is externally generated, and another stakeholder is relevant for its correctness.

  Regarding data sovereignty, it has to be mentioned that policy enforcement is still challenging but partly possible. Nevertheless, full control and self-determination are not possible for any of these approaches since a baseline of trust is always required in a decentralized network with multiple data providers and consumers and, therefore, for all approaches \cite{Scherenberg2024}.
  
  There are still some limitations and open topics in this field that could not be addressed by this paper to guide future research. For example, the rise in complexity of the sharing scenario where data is offered to a third party that has been provided by another party has to be evaluated regarding data sovereignty. Approach 3 mentions that the policies are copied in this case, but the actual handling of this scenario is still something to investigate. Mastering the complexity in the sense of finding best practices for data sovereignty and ownership, in general, is an important topic. This includes the need for stronger policy enforcement mechanisms and the ability to enable data sovereignty, even if data leaves the trusted environment and known applications. Also, the considered use cases and respective requirements need to be validated, specifically concerning the non-functional requirements, e.g., on response time, stemming from use cases such as feedback to design and regulatory demands in the form of Digital Product Passports.
  
  Additionally, it has to be highlighted that this publication is based on a German automotive research project and thus has a strong industry bias toward the automotive industry. The transfer to other industries is a topic that needs further investigation. In the automotive industry, the DT update approaches need to be scaled and validated in industrial environments as well. As a running network, Catena-X has the potential to enable this validation in the automotive sector.
  
\section{Conclusion}
\label{sec:conclusion}

In this paper, we pointed out what must be considered once a DT is updated across a network with different stakeholders, considering data sovereignty, digital continuity, and data ownership. We used an example from the Circular Economy use case of the research project of the German automotive industry Catena-X to show these considerations.

Three approaches to updating have been presented and elaborated upon regarding their strengths and weaknesses in relation to the above-mentioned marginal conditions. Further, open issues for all approaches have been shown. 

\section*{Acknowledgements}

The results of this publication are partly generated in the research and development project Catena-X, funded by the German Federal Ministry of Economics and Climate Protection (BMWK) under funding number 13IK004. 
We would like to thank the project participants for their input on the use cases and technical solutions.

\bibliography{literature.bib}
\bibliographystyle{ieeetr}

\end{document}

%% file: glossaries.tex
\newacronym{AAS}{AAS}{Asset Administration Shell}
\newacronym{ADR}{ADR}{Architecture Decision Record}
\newacronym{B2C}{B2C}{Business to Customer Relationship}
\newacronym{BoM}{BoM}{Bill of Material}
\newacronym{CE}{CE}{Circular Economy}
\newacronym{COTS}{COTS}{Commercials of the shelf applications}
\newacronym{DT}{DT}{Digital Twin}
\newacronym{EoL}{EoL}{End of life}
\newacronym{ESPR}{ESPR}{Ecodesign for Sustainable Products Regulation}
\newacronym{FBB}{FBB}{Functional Building Block}
\newacronym{FOSS}{FOSS}{Free and Open-Source Software}
\newacronym{SAFe}{SAFe}{Scaled Agile Framework}
\newacronym{VIN}{VIN}{Vehicle Identification Number}

%% file: abstract.tex
Digital Twins have evolved as a concept describing digital representations of physical assets. 
They can be used to facilitate simulations, monitoring, or optimization of product lifecycles.

Considering the concept of a Circular Economy, which entails several lifecycles of, e.g., vehicles, their components, and materials, it is important to investigate how the respective Digital Twins are managed over the lifecycle of their physical assets. 
This publication presents and compares three approaches for managing Digital Twins in industrial use cases. 
The analysis considers aspects such as updates, data ownership, and data sovereignty. 
The results based on the research project Catena-X

%% file: fig1_DTcreation.tex
\definecolor{plantucolor0000}{RGB}{0,0,0}
\definecolor{plantucolor0001}{RGB}{24,24,24}
\definecolor{plantucolor0002}{RGB}{226,226,240}
\definecolor{plantucolor0003}{RGB}{238,238,238}
\begin{tikzpicture}[yscale=-1
,pstyle0/.style={color=black,line width=1.5pt}
,pstyle1/.style={color=plantucolor0001,line width=0.5pt,dash pattern=on 5.0pt off 5.0pt}
,pstyle2/.style={color=plantucolor0001,fill=plantucolor0002,line width=0.5pt}
,pstyle3/.style={color=black,fill=plantucolor0003,line width=1.5pt}
,pstyle4/.style={color=plantucolor0001,line width=1.0pt}
,pstyle5/.style={color=plantucolor0001,fill=plantucolor0001,line width=1.0pt}
]
\draw[pstyle0] (10pt,56.0679pt) rectangle (375.8717pt,120.4799pt);
\draw[pstyle0] (10pt,241.598pt) rectangle (927.794pt,450.5401pt);
\draw[pstyle1] (100pt,39.0679pt) -- (100pt,628.7762pt);
\draw[pstyle1] (370.1717pt,39.0679pt) -- (370.1717pt,628.7762pt);
\draw[pstyle1] (630.3679pt,39.0679pt) -- (630.3679pt,628.7762pt);
\draw[pstyle1] (740.7326pt,39.0679pt) -- (740.7326pt,628.7762pt);
\draw[pstyle1] (922.6379pt,39.0679pt) -- (922.6379pt,628.7762pt);
\draw[pstyle2] (20pt,10pt) arc (180:270:5pt) -- (25pt,5pt) -- (176.1237pt,5pt) arc (270:360:5pt) -- (181.1237pt,10pt) -- (181.1237pt,33.0679pt) arc (0:90:5pt) -- (176.1237pt,38.0679pt) -- (25pt,38.0679pt) arc (90:180:5pt) -- (20pt,33.0679pt) -- cycle;
\node at (27pt,12pt)[below right,color=black]{Supplier A, B and C};
\draw[pstyle2] (20pt,632.7762pt) arc (180:270:5pt) -- (25pt,627.7762pt) -- (176.1237pt,627.7762pt) arc (270:360:5pt) -- (181.1237pt,632.7762pt) -- (181.1237pt,655.8441pt) arc (0:90:5pt) -- (176.1237pt,660.8441pt) -- (25pt,660.8441pt) arc (90:180:5pt) -- (20pt,655.8441pt) -- cycle;
\node at (27pt,634.7762pt)[below right,color=black]{Supplier A, B and C};
\draw[pstyle2] (345.1717pt,10pt) arc (180:270:5pt) -- (350.1717pt,5pt) -- (391.5717pt,5pt) arc (270:360:5pt) -- (396.5717pt,10pt) -- (396.5717pt,33.0679pt) arc (0:90:5pt) -- (391.5717pt,38.0679pt) -- (350.1717pt,38.0679pt) arc (90:180:5pt) -- (345.1717pt,33.0679pt) -- cycle;
\node at (352.1717pt,12pt)[below right,color=black]{OEM};
\draw[pstyle2] (345.1717pt,632.7762pt) arc (180:270:5pt) -- (350.1717pt,627.7762pt) -- (391.5717pt,627.7762pt) arc (270:360:5pt) -- (396.5717pt,632.7762pt) -- (396.5717pt,655.8441pt) arc (0:90:5pt) -- (391.5717pt,660.8441pt) -- (350.1717pt,660.8441pt) arc (90:180:5pt) -- (345.1717pt,655.8441pt) -- cycle;
\node at (352.1717pt,634.7762pt)[below right,color=black]{OEM};
\draw[pstyle2] (584.3679pt,10pt) arc (180:270:5pt) -- (589.3679pt,5pt) -- (671.7326pt,5pt) arc (270:360:5pt) -- (676.7326pt,10pt) -- (676.7326pt,33.0679pt) arc (0:90:5pt) -- (671.7326pt,38.0679pt) -- (589.3679pt,38.0679pt) arc (90:180:5pt) -- (584.3679pt,33.0679pt) -- cycle;
\node at (591.3679pt,12pt)[below right,color=black]{Consumer};
\draw[pstyle2] (584.3679pt,632.7762pt) arc (180:270:5pt) -- (589.3679pt,627.7762pt) -- (671.7326pt,627.7762pt) arc (270:360:5pt) -- (676.7326pt,632.7762pt) -- (676.7326pt,655.8441pt) arc (0:90:5pt) -- (671.7326pt,660.8441pt) -- (589.3679pt,660.8441pt) arc (90:180:5pt) -- (584.3679pt,655.8441pt) -- cycle;
\node at (591.3679pt,634.7762pt)[below right,color=black]{Consumer};
\draw[pstyle2] (686.7326pt,10pt) arc (180:270:5pt) -- (691.7326pt,5pt) -- (790.035pt,5pt) arc (270:360:5pt) -- (795.035pt,10pt) -- (795.035pt,33.0679pt) arc (0:90:5pt) -- (790.035pt,38.0679pt) -- (691.7326pt,38.0679pt) arc (90:180:5pt) -- (686.7326pt,33.0679pt) -- cycle;
\node at (693.7326pt,12pt)[below right,color=black]{Repair Shop};
\draw[pstyle2] (686.7326pt,632.7762pt) arc (180:270:5pt) -- (691.7326pt,627.7762pt) -- (790.035pt,627.7762pt) arc (270:360:5pt) -- (795.035pt,632.7762pt) -- (795.035pt,655.8441pt) arc (0:90:5pt) -- (790.035pt,660.8441pt) -- (691.7326pt,660.8441pt) arc (90:180:5pt) -- (686.7326pt,655.8441pt) -- cycle;
\node at (693.7326pt,634.7762pt)[below right,color=black]{Repair Shop};
\draw[pstyle2] (873.6379pt,10pt) arc (180:270:5pt) -- (878.6379pt,5pt) -- (966.9502pt,5pt) arc (270:360:5pt) -- (971.9502pt,10pt) -- (971.9502pt,33.0679pt) arc (0:90:5pt) -- (966.9502pt,38.0679pt) -- (878.6379pt,38.0679pt) arc (90:180:5pt) -- (873.6379pt,33.0679pt) -- cycle;
\node at (880.6379pt,12pt)[below right,color=black]{Dismantler};
\draw[pstyle2] (873.6379pt,632.7762pt) arc (180:270:5pt) -- (878.6379pt,627.7762pt) -- (966.9502pt,627.7762pt) arc (270:360:5pt) -- (971.9502pt,632.7762pt) -- (971.9502pt,655.8441pt) arc (0:90:5pt) -- (966.9502pt,660.8441pt) -- (878.6379pt,660.8441pt) arc (90:180:5pt) -- (873.6379pt,655.8441pt) -- cycle;
\node at (880.6379pt,634.7762pt)[below right,color=black]{Dismantler};
\draw[pstyle3] (10pt,56.0679pt) -- (89.199pt,56.0679pt) -- (89.199pt,65.7739pt) -- (79.199pt,75.7739pt) -- (10pt,75.7739pt) -- (10pt,56.0679pt);
\draw[pstyle0] (10pt,56.0679pt) rectangle (375.8717pt,120.4799pt);
\node at (25pt,57.0679pt)[below right,color=black]{\textbf{loop}};
\node at (104.199pt,58.0679pt)[below right,color=black]{\textbf{[for each produced component]}};
\draw[pstyle4] (100.5618pt,99.4799pt) -- (142.5618pt,99.4799pt);
\draw[pstyle4] (142.5618pt,99.4799pt) -- (142.5618pt,112.4799pt);
\draw[pstyle4] (101.5618pt,112.4799pt) -- (142.5618pt,112.4799pt);
\draw[pstyle5] (111.5618pt,108.4799pt) -- (101.5618pt,112.4799pt) -- (111.5618pt,116.4799pt) -- (107.5618pt,112.4799pt) -- cycle;
\node at (107.5618pt,79.7739pt)[below right,color=black]{Generate Digital Twin of component};
\draw[pstyle4] (370.8717pt,151.186pt) -- (412.8717pt,151.186pt);
\draw[pstyle4] (412.8717pt,151.186pt) -- (412.8717pt,164.186pt);
\draw[pstyle4] (371.8717pt,164.186pt) -- (412.8717pt,164.186pt);
\draw[pstyle5] (381.8717pt,160.186pt) -- (371.8717pt,164.186pt) -- (381.8717pt,168.186pt) -- (377.8717pt,164.186pt) -- cycle;
\node at (377.8717pt,131.4799pt)[below right,color=black]{Generate Digital Twin of vehicle};
\draw[pstyle4] (370.8717pt,213.598pt) -- (412.8717pt,213.598pt);
\draw[pstyle4] (412.8717pt,213.598pt) -- (412.8717pt,226.598pt);
\draw[pstyle4] (371.8717pt,226.598pt) -- (412.8717pt,226.598pt);
\draw[pstyle5] (381.8717pt,222.598pt) -- (371.8717pt,226.598pt) -- (381.8717pt,230.598pt) -- (377.8717pt,226.598pt) -- cycle;
\node at (377.8717pt,176.186pt)[below right,color=black]{Integrate Digital Twins of};
\node at (377.8717pt,193.892pt)[below right,color=black]{ components from supplier in BoM};
\draw[pstyle3] (10pt,241.598pt) -- (89.199pt,241.598pt) -- (89.199pt,251.304pt) -- (79.199pt,261.304pt) -- (10pt,261.304pt) -- (10pt,241.598pt);
\draw[pstyle0] (10pt,241.598pt) rectangle (927.794pt,450.5401pt);
\node at (25pt,242.598pt)[below right,color=black]{\textbf{loop}};
\node at (104.199pt,243.598pt)[below right,color=black]{\textbf{[whenever the vehicle is malfunctioning or components have to be maintained in Repair Shop]}};
\draw[pstyle5] (381.8717pt,281.01pt) -- (371.8717pt,285.01pt) -- (381.8717pt,289.01pt) -- (377.8717pt,285.01pt) -- cycle;
\draw[pstyle4] (375.8717pt,285.01pt) -- (739.8838pt,285.01pt);
\node at (387.8717pt,265.304pt)[below right,color=black]{request information on vehicle including BoM};
\draw[pstyle5] (111.5618pt,312.716pt) -- (101.5618pt,316.716pt) -- (111.5618pt,320.716pt) -- (107.5618pt,316.716pt) -- cycle;
\draw[pstyle4] (105.5618pt,316.716pt) -- (739.8838pt,316.716pt);
\node at (117.5618pt,297.01pt)[below right,color=black]{request information on component};
\draw[pstyle4] (740.8838pt,366.1281pt) -- (782.8838pt,366.1281pt);
\draw[pstyle4] (782.8838pt,366.1281pt) -- (782.8838pt,379.1281pt);
\draw[pstyle4] (741.8838pt,379.1281pt) -- (782.8838pt,379.1281pt);
\draw[pstyle5] (751.8838pt,375.1281pt) -- (741.8838pt,379.1281pt) -- (751.8838pt,383.1281pt) -- (747.8838pt,379.1281pt) -- cycle;
\node at (747.8838pt,328.716pt)[below right,color=black]{exchange component(s)};
\node at (747.8838pt,346.4221pt)[below right,color=black]{ or overhaul them};
\draw[pstyle5] (381.8717pt,406.8341pt) -- (371.8717pt,410.8341pt) -- (381.8717pt,414.8341pt) -- (377.8717pt,410.8341pt) -- cycle;
\draw[pstyle4] (375.8717pt,410.8341pt) -- (739.8838pt,410.8341pt);
\node at (387.8717pt,391.1281pt)[below right,color=black]{update scenario for vehicle};
\draw[pstyle5] (111.5618pt,438.5401pt) -- (101.5618pt,442.5401pt) -- (111.5618pt,446.5401pt) -- (107.5618pt,442.5401pt) -- cycle;
\draw[pstyle4] (105.5618pt,442.5401pt) -- (739.8838pt,442.5401pt);
\node at (117.5618pt,422.8341pt)[below right,color=black]{update scenario for components};
\draw[pstyle4] (922.794pt,534.3642pt) -- (964.794pt,534.3642pt);
\draw[pstyle4] (964.794pt,534.3642pt) -- (964.794pt,547.3642pt);
\draw[pstyle4] (923.794pt,547.3642pt) -- (964.794pt,547.3642pt);
\draw[pstyle5] (933.794pt,543.3642pt) -- (923.794pt,547.3642pt) -- (933.794pt,551.3642pt) -- (929.794pt,547.3642pt) -- cycle;
\node at (929.794pt,461.5401pt)[below right,color=black]{dismantle vehicle};
\node at (929.794pt,479.2461pt)[below right,color=black]{ and set state of decomissioning };
\node at (929.794pt,496.9521pt)[below right,color=black]{ for components and vehicle};
\node at (929.794pt,514.6582pt)[below right,color=black]{ };
\draw[pstyle5] (381.8717pt,575.0702pt) -- (371.8717pt,579.0702pt) -- (381.8717pt,583.0702pt) -- (377.8717pt,579.0702pt) -- cycle;
\draw[pstyle4] (375.8717pt,579.0702pt) -- (921.794pt,579.0702pt);
\node at (387.8717pt,559.3642pt)[below right,color=black]{update scenario for vehicle};
\draw[pstyle5] (111.5618pt,606.7762pt) -- (101.5618pt,610.7762pt) -- (111.5618pt,614.7762pt) -- (107.5618pt,610.7762pt) -- cycle;
\draw[pstyle4] (105.5618pt,610.7762pt) -- (921.794pt,610.7762pt);
\node at (117.5618pt,591.0702pt)[below right,color=black]{update scenario for components};
\end{tikzpicture}

%% file: fig3_approach1.tex
\definecolor{plantucolor0000}{RGB}{0,0,0}
\definecolor{plantucolor0001}{RGB}{24,24,24}
\definecolor{plantucolor0002}{RGB}{226,226,240}
\definecolor{plantucolor0003}{RGB}{238,238,238}
\begin{tikzpicture}[yscale=-1
,pstyle0/.style={color=black,line width=1.5pt}
,pstyle1/.style={color=plantucolor0001,line width=0.5pt,dash pattern=on 5.0pt off 5.0pt}
,pstyle2/.style={color=plantucolor0001,fill=plantucolor0002,line width=0.5pt}
,pstyle3/.style={color=plantucolor0001,fill=plantucolor0001,line width=1.0pt}
,pstyle4/.style={color=plantucolor0001,line width=1.0pt}
]
\draw[pstyle0] (10pt,87.7739pt) rectangle (635.7986pt,139.186pt);
\draw[pstyle1] (59pt,39.0679pt) -- (59pt,219.598pt);
\draw[pstyle1] (134.2421pt,39.0679pt) -- (134.2421pt,219.598pt);
\draw[pstyle1] (571.4962pt,39.0679pt) -- (571.4962pt,219.598pt);
\draw[pstyle1] (684.7986pt,39.0679pt) -- (684.7986pt,219.598pt);
\draw[pstyle2] (20pt,10pt) arc (180:270:5pt) -- (25pt,5pt) -- (94.2421pt,5pt) arc (270:360:5pt) -- (99.2421pt,10pt) -- (99.2421pt,33.0679pt) arc (0:90:5pt) -- (94.2421pt,38.0679pt) -- (25pt,38.0679pt) arc (90:180:5pt) -- (20pt,33.0679pt) -- cycle;
\node at (27pt,12pt)[below right,color=black]{Supplier};
\draw[pstyle2] (20pt,223.598pt) arc (180:270:5pt) -- (25pt,218.598pt) -- (94.2421pt,218.598pt) arc (270:360:5pt) -- (99.2421pt,223.598pt) -- (99.2421pt,246.6659pt) arc (0:90:5pt) -- (94.2421pt,251.6659pt) -- (25pt,251.6659pt) arc (90:180:5pt) -- (20pt,246.6659pt) -- cycle;
\node at (27pt,225.598pt)[below right,color=black]{Supplier};
\draw[pstyle2] (109.2421pt,10pt) arc (180:270:5pt) -- (114.2421pt,5pt) -- (155.6421pt,5pt) arc (270:360:5pt) -- (160.6421pt,10pt) -- (160.6421pt,33.0679pt) arc (0:90:5pt) -- (155.6421pt,38.0679pt) -- (114.2421pt,38.0679pt) arc (90:180:5pt) -- (109.2421pt,33.0679pt) -- cycle;
\node at (116.2421pt,12pt)[below right,color=black]{OEM};
\draw[pstyle2] (109.2421pt,223.598pt) arc (180:270:5pt) -- (114.2421pt,218.598pt) -- (155.6421pt,218.598pt) arc (270:360:5pt) -- (160.6421pt,223.598pt) -- (160.6421pt,246.6659pt) arc (0:90:5pt) -- (155.6421pt,251.6659pt) -- (114.2421pt,251.6659pt) arc (90:180:5pt) -- (109.2421pt,246.6659pt) -- cycle;
\node at (116.2421pt,225.598pt)[below right,color=black]{OEM};
\draw[pstyle2] (517.4962pt,10pt) arc (180:270:5pt) -- (522.4962pt,5pt) -- (620.7986pt,5pt) arc (270:360:5pt) -- (625.7986pt,10pt) -- (625.7986pt,33.0679pt) arc (0:90:5pt) -- (620.7986pt,38.0679pt) -- (522.4962pt,38.0679pt) arc (90:180:5pt) -- (517.4962pt,33.0679pt) -- cycle;
\node at (524.4962pt,12pt)[below right,color=black]{Repair Shop};
\draw[pstyle2] (517.4962pt,223.598pt) arc (180:270:5pt) -- (522.4962pt,218.598pt) -- (620.7986pt,218.598pt) arc (270:360:5pt) -- (625.7986pt,223.598pt) -- (625.7986pt,246.6659pt) arc (0:90:5pt) -- (620.7986pt,251.6659pt) -- (522.4962pt,251.6659pt) arc (90:180:5pt) -- (517.4962pt,246.6659pt) -- cycle;
\node at (524.4962pt,225.598pt)[below right,color=black]{Repair Shop};
\draw[pstyle2] (635.7986pt,10pt) arc (180:270:5pt) -- (640.7986pt,5pt) -- (729.1109pt,5pt) arc (270:360:5pt) -- (734.1109pt,10pt) -- (734.1109pt,33.0679pt) arc (0:90:5pt) -- (729.1109pt,38.0679pt) -- (640.7986pt,38.0679pt) arc (90:180:5pt) -- (635.7986pt,33.0679pt) -- cycle;
\node at (642.7986pt,12pt)[below right,color=black]{Dismantler};
\draw[pstyle2] (635.7986pt,223.598pt) arc (180:270:5pt) -- (640.7986pt,218.598pt) -- (729.1109pt,218.598pt) arc (270:360:5pt) -- (734.1109pt,223.598pt) -- (734.1109pt,246.6659pt) arc (0:90:5pt) -- (729.1109pt,251.6659pt) -- (640.7986pt,251.6659pt) arc (90:180:5pt) -- (635.7986pt,246.6659pt) -- cycle;
\node at (642.7986pt,225.598pt)[below right,color=black]{Dismantler};
\draw[pstyle3] (145.9421pt,68.7739pt) -- (135.9421pt,72.7739pt) -- (145.9421pt,76.7739pt) -- (141.9421pt,72.7739pt) -- cycle;
\draw[pstyle4] (139.9421pt,72.7739pt) -- (570.6474pt,72.7739pt);
\node at (151.9421pt,53.0679pt)[below right,color=black]{write updates on vehicle into Digital Twin Registry of OEM};
\draw[color=black,fill=plantucolor0003,line width=1.5pt] (10pt,87.7739pt) -- (80.838pt,87.7739pt) -- (80.838pt,97.4799pt) -- (70.838pt,107.4799pt) -- (10pt,107.4799pt) -- (10pt,87.7739pt);
\draw[pstyle0] (10pt,87.7739pt) rectangle (635.7986pt,139.186pt);
\node at (25pt,88.7739pt)[below right,color=black]{\textbf{opt}};
\node at (95.838pt,89.7739pt)[below right,color=black]{\textbf{[when component is overhauled or otherwise modified]}};
\draw[pstyle3] (70.6211pt,127.186pt) -- (60.6211pt,131.186pt) -- (70.6211pt,135.186pt) -- (66.6211pt,131.186pt) -- cycle;
\draw[pstyle4] (64.6211pt,131.186pt) -- (570.6474pt,131.186pt);
\node at (76.6211pt,111.4799pt)[below right,color=black]{write updates on component into Digital Twin Registry of supplier};
\draw[pstyle3] (145.9421pt,165.892pt) -- (135.9421pt,169.892pt) -- (145.9421pt,173.892pt) -- (141.9421pt,169.892pt) -- cycle;
\draw[pstyle4] (139.9421pt,169.892pt) -- (683.9548pt,169.892pt);
\node at (151.9421pt,150.186pt)[below right,color=black]{write updates on vehicle into Digital Twin Registry of OEM};
\draw[pstyle3] (70.6211pt,197.598pt) -- (60.6211pt,201.598pt) -- (70.6211pt,205.598pt) -- (66.6211pt,201.598pt) -- cycle;
\draw[pstyle4] (64.6211pt,201.598pt) -- (683.9548pt,201.598pt);
\node at (76.6211pt,181.892pt)[below right,color=black]{write updates on component into Digital Twin Registry of supplier};
\end{tikzpicture}

%% file: fig4_approach2.tex
\definecolor{plantucolor0000}{RGB}{0,0,0}
\definecolor{plantucolor0001}{RGB}{24,24,24}
\definecolor{plantucolor0002}{RGB}{226,226,240}
\definecolor{plantucolor0003}{RGB}{238,238,238}
\begin{tikzpicture}[yscale=-1
,pstyle0/.style={color=black,line width=1.5pt}
,pstyle1/.style={color=plantucolor0001,line width=0.5pt,dash pattern=on 5.0pt off 5.0pt}
,pstyle2/.style={color=plantucolor0001,fill=plantucolor0002,line width=0.5pt}
,pstyle4/.style={color=plantucolor0001,line width=1.0pt}
,pstyle5/.style={color=plantucolor0001,fill=plantucolor0001,line width=1.0pt}
]
\draw[pstyle0] (145.6421pt,56.0679pt) rectangle (550.5578pt,167.186pt);
\draw[pstyle1] (44pt,39.0679pt) -- (44pt,228.892pt);
\draw[pstyle1] (119.2421pt,39.0679pt) -- (119.2421pt,228.892pt);
\draw[pstyle1] (209.6421pt,39.0679pt) -- (209.6421pt,228.892pt);
\draw[pstyle1] (504.4388pt,39.0679pt) -- (504.4388pt,228.892pt);
\draw[pstyle2] (5pt,10pt) arc (180:270:5pt) -- (10pt,5pt) -- (79.2421pt,5pt) arc (270:360:5pt) -- (84.2421pt,10pt) -- (84.2421pt,33.0679pt) arc (0:90:5pt) -- (79.2421pt,38.0679pt) -- (10pt,38.0679pt) arc (90:180:5pt) -- (5pt,33.0679pt) -- cycle;
\node at (12pt,12pt)[below right,color=black]{Supplier};
\draw[pstyle2] (5pt,232.892pt) arc (180:270:5pt) -- (10pt,227.892pt) -- (79.2421pt,227.892pt) arc (270:360:5pt) -- (84.2421pt,232.892pt) -- (84.2421pt,255.9599pt) arc (0:90:5pt) -- (79.2421pt,260.9599pt) -- (10pt,260.9599pt) arc (90:180:5pt) -- (5pt,255.9599pt) -- cycle;
\node at (12pt,234.892pt)[below right,color=black]{Supplier};
\draw[pstyle2] (94.2421pt,10pt) arc (180:270:5pt) -- (99.2421pt,5pt) -- (140.6421pt,5pt) arc (270:360:5pt) -- (145.6421pt,10pt) -- (145.6421pt,33.0679pt) arc (0:90:5pt) -- (140.6421pt,38.0679pt) -- (99.2421pt,38.0679pt) arc (90:180:5pt) -- (94.2421pt,33.0679pt) -- cycle;
\node at (101.2421pt,12pt)[below right,color=black]{OEM};
\draw[pstyle2] (94.2421pt,232.892pt) arc (180:270:5pt) -- (99.2421pt,227.892pt) -- (140.6421pt,227.892pt) arc (270:360:5pt) -- (145.6421pt,232.892pt) -- (145.6421pt,255.9599pt) arc (0:90:5pt) -- (140.6421pt,260.9599pt) -- (99.2421pt,260.9599pt) arc (90:180:5pt) -- (94.2421pt,255.9599pt) -- cycle;
\node at (101.2421pt,234.892pt)[below right,color=black]{OEM};
\draw[pstyle2] (155.6421pt,10pt) arc (180:270:5pt) -- (160.6421pt,5pt) -- (258.9445pt,5pt) arc (270:360:5pt) -- (263.9445pt,10pt) -- (263.9445pt,33.0679pt) arc (0:90:5pt) -- (258.9445pt,38.0679pt) -- (160.6421pt,38.0679pt) arc (90:180:5pt) -- (155.6421pt,33.0679pt) -- cycle;
\node at (162.6421pt,12pt)[below right,color=black]{Repair Shop};
\draw[pstyle2] (155.6421pt,232.892pt) arc (180:270:5pt) -- (160.6421pt,227.892pt) -- (258.9445pt,227.892pt) arc (270:360:5pt) -- (263.9445pt,232.892pt) -- (263.9445pt,255.9599pt) arc (0:90:5pt) -- (258.9445pt,260.9599pt) -- (160.6421pt,260.9599pt) arc (90:180:5pt) -- (155.6421pt,255.9599pt) -- cycle;
\node at (162.6421pt,234.892pt)[below right,color=black]{Repair Shop};
\draw[pstyle2] (455.4388pt,10pt) arc (180:270:5pt) -- (460.4388pt,5pt) -- (548.7511pt,5pt) arc (270:360:5pt) -- (553.7511pt,10pt) -- (553.7511pt,33.0679pt) arc (0:90:5pt) -- (548.7511pt,38.0679pt) -- (460.4388pt,38.0679pt) arc (90:180:5pt) -- (455.4388pt,33.0679pt) -- cycle;
\node at (462.4388pt,12pt)[below right,color=black]{Dismantler};
\draw[pstyle2] (455.4388pt,232.892pt) arc (180:270:5pt) -- (460.4388pt,227.892pt) -- (548.7511pt,227.892pt) arc (270:360:5pt) -- (553.7511pt,232.892pt) -- (553.7511pt,255.9599pt) arc (0:90:5pt) -- (548.7511pt,260.9599pt) -- (460.4388pt,260.9599pt) arc (90:180:5pt) -- (455.4388pt,255.9599pt) -- cycle;
\node at (462.4388pt,234.892pt)[below right,color=black]{Dismantler};
\draw[color=black,fill=plantucolor0003,line width=1.5pt] (145.6421pt,56.0679pt) -- (212.3156pt,56.0679pt) -- (212.3156pt,65.7739pt) -- (202.3156pt,75.7739pt) -- (145.6421pt,75.7739pt) -- (145.6421pt,56.0679pt);
\draw[pstyle0] (145.6421pt,56.0679pt) rectangle (550.5578pt,167.186pt);
\node at (160.6421pt,57.0679pt)[below right,color=black]{\textbf{alt}};
\node at (227.3156pt,58.0679pt)[below right,color=black]{\textbf{[Digital Twin does not yet exist at Repair Shop]}};
\draw[pstyle4] (209.7933pt,99.4799pt) -- (251.7933pt,99.4799pt);
\draw[pstyle4] (251.7933pt,99.4799pt) -- (251.7933pt,112.4799pt);
\draw[pstyle4] (210.7933pt,112.4799pt) -- (251.7933pt,112.4799pt);
\draw[pstyle5] (220.7933pt,108.4799pt) -- (210.7933pt,112.4799pt) -- (220.7933pt,116.4799pt) -- (216.7933pt,112.4799pt) -- cycle;
\node at (216.7933pt,79.7739pt)[below right,color=black]{create own Digital Twin at Repair Shop};
\draw[color=black,line width=1.0pt,dash pattern=on 2.0pt off 2.0pt] (145.6421pt,121.4799pt) -- (550.5578pt,121.4799pt);
\draw[pstyle4] (209.7933pt,146.186pt) -- (251.7933pt,146.186pt);
\draw[pstyle4] (251.7933pt,146.186pt) -- (251.7933pt,159.186pt);
\draw[pstyle4] (210.7933pt,159.186pt) -- (251.7933pt,159.186pt);
\draw[pstyle5] (220.7933pt,155.186pt) -- (210.7933pt,159.186pt) -- (220.7933pt,163.186pt) -- (216.7933pt,159.186pt) -- cycle;
\node at (216.7933pt,126.4799pt)[below right,color=black]{update own Digital Twin at Repair Shop};
\draw[pstyle4] (504.595pt,197.892pt) -- (546.595pt,197.892pt);
\draw[pstyle4] (546.595pt,197.892pt) -- (546.595pt,210.892pt);
\draw[pstyle4] (505.595pt,210.892pt) -- (546.595pt,210.892pt);
\draw[pstyle5] (515.595pt,206.892pt) -- (505.595pt,210.892pt) -- (515.595pt,214.892pt) -- (511.595pt,210.892pt) -- cycle;
\node at (511.595pt,178.186pt)[below right,color=black]{create own Digital Twin at Dismantler};
\end{tikzpicture}

%% file: fig5_approach3.tex
\definecolor{plantucolor0000}{RGB}{0,0,0}
\definecolor{plantucolor0001}{RGB}{24,24,24}
\definecolor{plantucolor0002}{RGB}{226,226,240}
\definecolor{plantucolor0003}{RGB}{238,238,238}
\begin{tikzpicture}[yscale=-1
,pstyle0/.style={color=black,line width=1.5pt}
,pstyle1/.style={color=plantucolor0001,line width=0.5pt,dash pattern=on 5.0pt off 5.0pt}
,pstyle2/.style={color=plantucolor0001,fill=plantucolor0002,line width=0.5pt}
,pstyle4/.style={color=plantucolor0001,line width=1.0pt}
,pstyle5/.style={color=plantucolor0001,fill=plantucolor0001,line width=1.0pt}
]
\draw[pstyle0] (84.2421pt,56.0679pt) rectangle (775.8657pt,269.716pt);
\draw[pstyle1] (44pt,39.0679pt) -- (44pt,578.4822pt);
\draw[pstyle1] (119.2421pt,39.0679pt) -- (119.2421pt,578.4822pt);
\draw[pstyle1] (649.2435pt,39.0679pt) -- (649.2435pt,578.4822pt);
\draw[pstyle1] (770.7095pt,39.0679pt) -- (770.7095pt,578.4822pt);
\draw[pstyle2] (5pt,10pt) arc (180:270:5pt) -- (10pt,5pt) -- (79.2421pt,5pt) arc (270:360:5pt) -- (84.2421pt,10pt) -- (84.2421pt,33.0679pt) arc (0:90:5pt) -- (79.2421pt,38.0679pt) -- (10pt,38.0679pt) arc (90:180:5pt) -- (5pt,33.0679pt) -- cycle;
\node at (12pt,12pt)[below right,color=black]{Supplier};
\draw[pstyle2] (5pt,582.4822pt) arc (180:270:5pt) -- (10pt,577.4822pt) -- (79.2421pt,577.4822pt) arc (270:360:5pt) -- (84.2421pt,582.4822pt) -- (84.2421pt,605.5501pt) arc (0:90:5pt) -- (79.2421pt,610.5501pt) -- (10pt,610.5501pt) arc (90:180:5pt) -- (5pt,605.5501pt) -- cycle;
\node at (12pt,584.4822pt)[below right,color=black]{Supplier};
\draw[pstyle2] (94.2421pt,10pt) arc (180:270:5pt) -- (99.2421pt,5pt) -- (140.6421pt,5pt) arc (270:360:5pt) -- (145.6421pt,10pt) -- (145.6421pt,33.0679pt) arc (0:90:5pt) -- (140.6421pt,38.0679pt) -- (99.2421pt,38.0679pt) arc (90:180:5pt) -- (94.2421pt,33.0679pt) -- cycle;
\node at (101.2421pt,12pt)[below right,color=black]{OEM};
\draw[pstyle2] (94.2421pt,582.4822pt) arc (180:270:5pt) -- (99.2421pt,577.4822pt) -- (140.6421pt,577.4822pt) arc (270:360:5pt) -- (145.6421pt,582.4822pt) -- (145.6421pt,605.5501pt) arc (0:90:5pt) -- (140.6421pt,610.5501pt) -- (99.2421pt,610.5501pt) arc (90:180:5pt) -- (94.2421pt,605.5501pt) -- cycle;
\node at (101.2421pt,584.4822pt)[below right,color=black]{OEM};
\draw[pstyle2] (595.2435pt,10pt) arc (180:270:5pt) -- (600.2435pt,5pt) -- (698.546pt,5pt) arc (270:360:5pt) -- (703.546pt,10pt) -- (703.546pt,33.0679pt) arc (0:90:5pt) -- (698.546pt,38.0679pt) -- (600.2435pt,38.0679pt) arc (90:180:5pt) -- (595.2435pt,33.0679pt) -- cycle;
\node at (602.2435pt,12pt)[below right,color=black]{Repair Shop};
\draw[pstyle2] (595.2435pt,582.4822pt) arc (180:270:5pt) -- (600.2435pt,577.4822pt) -- (698.546pt,577.4822pt) arc (270:360:5pt) -- (703.546pt,582.4822pt) -- (703.546pt,605.5501pt) arc (0:90:5pt) -- (698.546pt,610.5501pt) -- (600.2435pt,610.5501pt) arc (90:180:5pt) -- (595.2435pt,605.5501pt) -- cycle;
\node at (602.2435pt,584.4822pt)[below right,color=black]{Repair Shop};
\draw[pstyle2] (721.7095pt,10pt) arc (180:270:5pt) -- (726.7095pt,5pt) -- (815.0219pt,5pt) arc (270:360:5pt) -- (820.0219pt,10pt) -- (820.0219pt,33.0679pt) arc (0:90:5pt) -- (815.0219pt,38.0679pt) -- (726.7095pt,38.0679pt) arc (90:180:5pt) -- (721.7095pt,33.0679pt) -- cycle;
\node at (728.7095pt,12pt)[below right,color=black]{Dismantler};
\draw[pstyle2] (721.7095pt,582.4822pt) arc (180:270:5pt) -- (726.7095pt,577.4822pt) -- (815.0219pt,577.4822pt) arc (270:360:5pt) -- (820.0219pt,582.4822pt) -- (820.0219pt,605.5501pt) arc (0:90:5pt) -- (815.0219pt,610.5501pt) -- (726.7095pt,610.5501pt) arc (90:180:5pt) -- (721.7095pt,605.5501pt) -- cycle;
\node at (728.7095pt,584.4822pt)[below right,color=black]{Dismantler};
\draw[color=black,fill=plantucolor0003,line width=1.5pt] (84.2421pt,56.0679pt) -- (150.9156pt,56.0679pt) -- (150.9156pt,65.7739pt) -- (140.9156pt,75.7739pt) -- (84.2421pt,75.7739pt) -- (84.2421pt,56.0679pt);
\draw[pstyle0] (84.2421pt,56.0679pt) rectangle (775.8657pt,269.716pt);
\node at (99.2421pt,57.0679pt)[below right,color=black]{\textbf{alt}};
\node at (165.9156pt,58.0679pt)[below right,color=black]{\textbf{[Digital Twin does not yet exist at repair shop]}};
\draw[pstyle4] (649.3947pt,134.892pt) -- (691.3947pt,134.892pt);
\draw[pstyle4] (691.3947pt,134.892pt) -- (691.3947pt,147.892pt);
\draw[pstyle4] (650.3947pt,147.892pt) -- (691.3947pt,147.892pt);
\draw[pstyle5] (660.3947pt,143.892pt) -- (650.3947pt,147.892pt) -- (660.3947pt,151.892pt) -- (656.3947pt,147.892pt) -- cycle;
\node at (656.3947pt,79.7739pt)[below right,color=black]{create own};
\node at (656.3947pt,97.4799pt)[below right,color=black]{ Digital Twin};
\node at (656.3947pt,115.186pt)[below right,color=black]{ at Repair Shop};
\draw[pstyle5] (130.9421pt,175.598pt) -- (120.9421pt,179.598pt) -- (130.9421pt,183.598pt) -- (126.9421pt,179.598pt) -- cycle;
\draw[pstyle4] (124.9421pt,179.598pt) -- (648.3947pt,179.598pt);
\node at (136.9421pt,159.892pt)[below right,color=black]{request relevant accessible information and copy into own Digital Twin};
\draw[color=black,line width=1.0pt,dash pattern=on 2.0pt off 2.0pt] (84.2421pt,188.598pt) -- (775.8657pt,188.598pt);
\draw[pstyle4] (649.3947pt,248.716pt) -- (691.3947pt,248.716pt);
\draw[pstyle4] (691.3947pt,248.716pt) -- (691.3947pt,261.716pt);
\draw[pstyle4] (650.3947pt,261.716pt) -- (691.3947pt,261.716pt);
\draw[pstyle5] (660.3947pt,257.716pt) -- (650.3947pt,261.716pt) -- (660.3947pt,265.716pt) -- (656.3947pt,261.716pt) -- cycle;
\node at (656.3947pt,193.598pt)[below right,color=black]{update own};
\node at (656.3947pt,211.304pt)[below right,color=black]{ Digital Twin};
\node at (656.3947pt,229.01pt)[below right,color=black]{ at Repair Shop};
\draw[pstyle5] (130.9421pt,314.1281pt) -- (120.9421pt,318.1281pt) -- (130.9421pt,322.1281pt) -- (126.9421pt,318.1281pt) -- cycle;
\draw[pstyle4] (124.9421pt,318.1281pt) -- (648.3947pt,318.1281pt);
\node at (136.9421pt,280.716pt)[below right,color=black]{send a notification, that there have been updates};
\node at (136.9421pt,298.4221pt)[below right,color=black]{ and create an automatic notification};
\draw[pstyle5] (637.3947pt,363.5401pt) -- (647.3947pt,367.5401pt) -- (637.3947pt,371.5401pt) -- (641.3947pt,367.5401pt) -- cycle;
\draw[pstyle4] (119.9421pt,367.5401pt) -- (643.3947pt,367.5401pt);
\node at (126.9421pt,330.1281pt)[below right,color=black]{request, pull and integrate};
\node at (126.9421pt,347.8341pt)[below right,color=black]{ information into own vehicle Digital Twin};
\draw[pstyle4] (770.8657pt,434.6582pt) -- (812.8657pt,434.6582pt);
\draw[pstyle4] (812.8657pt,434.6582pt) -- (812.8657pt,447.6582pt);
\draw[pstyle4] (771.8657pt,447.6582pt) -- (812.8657pt,447.6582pt);
\draw[pstyle5] (781.8657pt,443.6582pt) -- (771.8657pt,447.6582pt) -- (781.8657pt,451.6582pt) -- (777.8657pt,447.6582pt) -- cycle;
\node at (777.8657pt,379.5401pt)[below right,color=black]{create own};
\node at (777.8657pt,397.2461pt)[below right,color=black]{ Digital Twin};
\node at (777.8657pt,414.9521pt)[below right,color=black]{ at Dismantler};
\draw[pstyle5] (758.8657pt,475.3642pt) -- (768.8657pt,479.3642pt) -- (758.8657pt,483.3642pt) -- (762.8657pt,479.3642pt) -- cycle;
\draw[pstyle4] (119.9421pt,479.3642pt) -- (764.8657pt,479.3642pt);
\node at (126.9421pt,459.6582pt)[below right,color=black]{copy relevant accessible information into own Digital Twin};
\draw[pstyle5] (130.9421pt,507.0702pt) -- (120.9421pt,511.0702pt) -- (130.9421pt,515.0702pt) -- (126.9421pt,511.0702pt) -- cycle;
\draw[pstyle4] (124.9421pt,511.0702pt) -- (769.8657pt,511.0702pt);
\node at (136.9421pt,491.3642pt)[below right,color=black]{send a notification, that there have been updates and create an automatic notification};
\draw[pstyle5] (758.8657pt,556.4822pt) -- (768.8657pt,560.4822pt) -- (758.8657pt,564.4822pt) -- (762.8657pt,560.4822pt) -- cycle;
\draw[pstyle4] (119.9421pt,560.4822pt) -- (764.8657pt,560.4822pt);
\node at (126.9421pt,523.0702pt)[below right,color=black]{request, pull and integrate};
\node at (126.9421pt,540.7762pt)[below right,color=black]{ information into own vehicle Digital Twin};
\end{tikzpicture}